\documentclass[twocolumn,aps,prl,showpacs,preprintnumbers]{revtex4-1}%
\usepackage{mathrsfs}
\usepackage{amsmath}
\usepackage{amsfonts}
\usepackage{amssymb}
\usepackage{amsthm}
\usepackage{graphicx}
\usepackage{color}
\usepackage{txfonts}
\usepackage{eucal}
\usepackage{array}
\usepackage{pifont}%
\usepackage[colorlinks=true,citecolor=blue,linkcolor=blue,urlcolor=blue,anchorcolor=blue]{hyperref}%
\hypersetup{colorlinks=true,citecolor=blue,linkcolor=blue,urlcolor=blue}
\begin{document}
\title{Exchange magnon-polaritons in microwave cavities}
\author{Yunshan Cao$^{1}$}
\author{Peng Yan$^{1}$}
\author{Hans Huebl$^{2,3,4}$}
\author{Sebastian T.B. Goennenwein$^{2,3,4}$}
\author{Gerrit E.W. Bauer$^{5,1}$}
\affiliation{$^{1}$Kavli Institute of NanoScience, Delft University of Technology,
Lorentzweg 1, 2628 CJ Delft, The Netherlands}
\affiliation{$^{2}$Walther-Mei{\ss }ner-Institute, Bayerische Akademie der Wissenschaften,
85748 Garching, Germany}
\affiliation{$^{3}$Nanosystems Initiative Munich, D-80799 M\"{u}nchen, Germany}
\affiliation{$^{4}$Physik-Department, Technische Universit\"{a}t M\"{u}nchen, D-85748 Garching, Germany}
\affiliation{$^{5}$Institute for Materials Research and WPI-AIMR, Tohoku University, Sendai
980-8577, Japan}

\begin{abstract}
We formulate a scattering theory to study magnetic films in microwave cavities
beyond the independent-spin and rotating wave approximations of the
Tavis-Cummings model. We demonstrate that strong coupling can be realized not
only for the ferromagnetic resonance (FMR) mode, but also for spin wave
resonances (SWRs); the coupling strengths are mode dependent and decrease with
increasing mode index. The strong coupling regime can be also accessed
electrically by spin pumping into a metal contact.

\end{abstract}

\date{\today}
\maketitle

\begin{center}
\textbf{I. INTRODUCTION}
\end{center}
Strong light-matter interaction is a central subject in quantum information
and communication science and technology.
Hybrid systems consisting of resonantly coupled spin ensembles and microwaves
received much attention recently \cite{Kubo2010, Soykal2010, Putz2014}.
In magnetic materials, spins are coupled by the exchange interactions into ordered states.
The collective elementary excitations of the spin system are spin waves or magnons.
Arguably the most important experimental technique
is the microwave spectroscopy of the magnetic order parameter called ferromagnetic resonance
(FMR) and/or spin wave resonance (SWR) \cite{Hillebrands},
which is usually used to study magnetism in the weak coupling limit.
In the strong coupling limit, the hybridized states of the magnetic order parameter with electromagnetic waves
are magnon-polaritons \cite{Mills1974, Lehmeyer1985}. They can be observed only when
the viscous damping of the magnetization dynamics
as parameterized by the Gilbert constant is sufficiently weak.
Of special interest from a materials perspective is yttrium iron garnet (YIG) \cite{Wu2013, Serga2010},
a ferrimagnetic insulator.
YIG is advantageous due to
(i) an extremely low dissipation, with Gilbert damping factor $\alpha$
down to $\sim10^{-5}$ \cite{Kajiwara2010};
(ii) a large spin density $2\times10^{22}~\mathrm{cm}^{-3}$ \cite{Gilleo1958}, much higher than in
paramagnetic materials which only have about $10^{15}\sim 10^{18}~\mathrm{cm}^{-3}$ \cite{Schuster2010, Abe2011}.
Therefore, strong coupling is much easier to achieve using YIG,
in either broad-band coplanar waveguides (CPWs) \cite{Huebl2013, Stenning2013, Bhoi2014}
or metallic microwave cavities \cite{Tabuchi2014, Zhang2014, Goryachev2014}.



\begin{figure}[tbh]
\centering
\includegraphics[width=8.5cm]{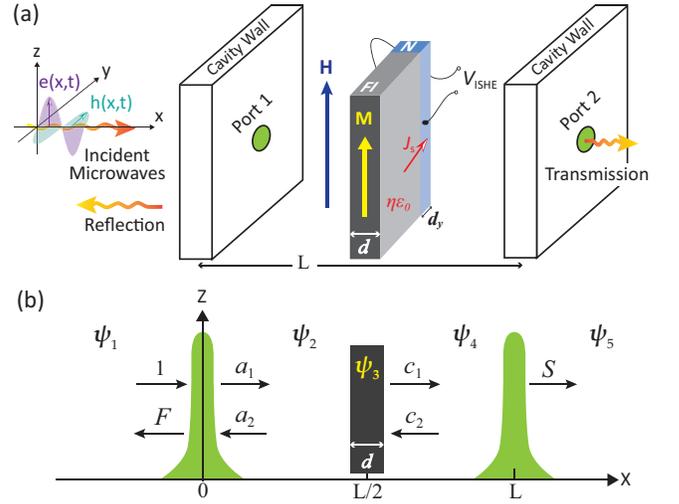}\caption{Magnetic film in a planar
microwave cavity.}%
\label{Fig1}%
\end{figure}

The conventional description for the coherent interaction between spins and photons
is based on the Tavis-Cummings (TC) model \cite{Fink2009}, where the effective coupling
strength $g_{\mathrm{eff}}=\sqrt{N}g_{s}$ of a single magnon
($N$ spins) to a single photon is enhanced by
$\sqrt{N}$ as compared to the coupling $g_{s}$ to a single spin. A standard
input-output formalism in the low photon number limit \cite{Milburn2008, Chiorescu2010}
provides the transmission amplitude
of microwaves from the input to the output port of the microwave resonator
(sketched in Fig. \ref{Fig1}(a)),
\begin{equation}
    S_{21}=\frac{\kappa_{\mathrm{e}}}{i(\omega-\omega_{\mathrm{c}})-(\kappa
_{\mathrm{e}}+\kappa_{\mathrm{i}})+\Sigma(\omega)}, \label{Eq-S21}%
\end{equation}
where $\omega_{\mathrm{c}},\kappa_{\mathrm{e,i}}$ are, respectively, the
resonance frequency and external/intrisic loss rates of the microwave
resonator (total damping rate $\kappa_{\mathrm{c}}=\kappa_{\mathrm{e}}%
+\kappa_{\mathrm{i}}$). The self-energy caused by the magnon-photon coupling
reads $\Sigma(\omega)=g_{\mathrm{eff}}^{2}/[i(\omega-\omega_{\mathrm{FMR}%
})-\kappa_{\mathrm{s}}]$, with FMR frequency $\omega_{\mathrm{FMR}}$ and
magnetic relaxation rate $\kappa_{\mathrm{s}}$. When $g_{\mathrm{eff}}%
>\kappa_{\mathrm{s,c}}$, the strong coupling regime is achieved
and explained well by the TC model
\cite{Huebl2013, Stenning2013, Bhoi2014, Tabuchi2014, Zhang2014, Goryachev2014, Fink2009, Schuster2010, Abe2011}.
However, the TC model based on monochrome mode interaction and
the rotating-wave approximation (RWA), fails to
describe the ultra-strong coupling (USC) regime and multi-mode behavior.
Although the TC model can in principle be repaired to cover the USC regime \cite{Agarwal2012},
the cited experiments investigated ferromagnetic samples of different shapes exposed to microwaves
in different geometries, which is beyond a generic TC model.
In this paper we present a first-principles theory that supersedes the TC model in treating
ferromagnetic objects coherently interacting with microwaves.


Huebl \textit{et al.} \cite{Huebl2013} demonstrated strong
coupling of a YIG film in a superconducting CPW in terms of
an anti-crossing in the microwave transmission spectrum when the
FMR matches the CPW frequency. A series of anti-crossings for thicker YIG samples indicative of spin wave
excitations are reported in YIG-film split-rings
\cite{Stenning2013, Bhoi2014}. Tabuchi \textit{et al.} \cite{Tabuchi2014} studied the strong
coupling regime for YIG spheres in 3D cavity system down to low temperatures
and subsequently coupled the magnon to a qubit via the microwave cavity mode.
Characteristic phenomena associated with distinct
parameter regimes, like magnetically induced transparency ($\kappa
_{s}<g_{\mathrm{eff}}<\kappa_{c}$) and Purcell effect ($\kappa_{c}%
<g_{\mathrm{eff}}<\kappa_{s}$), even the USC regime beyond the RWA were observed by Zhang \textit{et
al.} \cite{Zhang2014}. Goryachev \textit{et al.} \cite{Goryachev2014}
reported strong coupling between multiple magnon modes and a dark cavity mode
for submillimeter-size YIG spheres in 3D reentrant cavities,
as well as a high cooperativity of $> 10^5$ by USC to a bright cavity mode.

Strongly hybridized magnon-polaritons as observed in the above experiments
cannot be described in terms of a single magnon-photon coupling process.
In the present work, we formulate the coupling of a magnetic film to microwaves
in a cavity by means of scattering approach. Our method is valid for the full parameter range spanning
the weak to strong, even ultra-strong coupling limits. We obtain a general
transmission formula that reduces to the TC model in the appropriate limits.
To this end we solve the coupled Maxwell's and Landau-Lifshitz-Gilbert (LLG)
equations without making the conventional magnetostatic approximation. We may
then compute microwave absorption and transmission spectra that can be
characterized by multi-mode strong coupling and the mode-dependent coupling strengths.
Furthermore, we consider the electric detection in the strong coupling regime
through spin pumping \cite{Tserkovnyak} technique as measured in a Pt contact
by the inverse spin Hall effect (ISHE) \cite{Sandweg2010, Hoffmann2013}.

This paper is organized as follows: In Sec. II, we model the cavity and derive the equations of motion for coupled magnons and photons. Section III gives the formulation of the scattering theory and the main results of the magnon-photon strong coupling in both paramagnets and ferromagnets. An electric detection of the strong coupling is also proposed via spin pumping and inverse spin Hall effects. Conclusions are drawn in Sec. IV.

\begin{center}
\textbf{II. MODEL}
\end{center}

The weak to strong coupling transition can best be studied in a
simple configuration as shown in Fig. \ref{Fig1}(a). The calculations for
general configurations will be reported elsewhere. The magnetic film lies in
the $y$-$z$ plane between the cavity defining mirrors. The equilibrium
magnetization points into the $z$-direction by crystal anisotropy, dipolar,
and external magnetic fields. The incident microwave propagates along $x$ with
rf magnetic field linearly polarized along $y$. The cavity walls are modeled
by the permeability $\mu(x)=\mu_{0}\left[  1+2\ell
\delta(x)+2\ell\delta(x-L)\right]  $, where $L$ is the cavity width and $\ell$
models the wall opacity. In the absence of sources, the microwaves satisfy the
Maxwell's equation in frequency space,%
\begin{equation}
\partial_{x}^{2}\mathbf{h}(x)+\frac{\mu(x)}{\mu_{0}}%
q^{2}\mathbf{h}(x)=0,
\end{equation}
where $q=\omega/c$, with vacuum speed of light $c=1/\sqrt{\varepsilon_{0}%
\mu_{0}}$, and $\varepsilon_{0},\mu_{0}$ are the vacuum permittivity and
permeability, respectively.

Inside the magnetic film, we consider small-amplitude spatiotemporal
magnetizations $\mathbf{M}=M_{\mathrm{s}}\hat{z}+\mathbf{m}$, where
$M_{\mathrm{s}}$ is the saturation magnetization and $\mathbf{m}$ is driven by
the rf magnetic field $\mathbf{h}$, according to the Maxwell's equation
\begin{equation}
\left(  \nabla^{2}+k_{\varepsilon}^{2}\right)  \mathbf{h}(x)=\nabla
(\nabla\cdot\mathbf{h}(x))-k_{\varepsilon}^{2}\mathbf{m}(x), \label{Eq-Maxs}%
\end{equation}
where $\varepsilon$ is the permittivity of the magnet, $k_{\varepsilon}%
^{2}\equiv\varepsilon\mu_{0}\omega^{2}=\eta q^{2}$, and dielectric constant
$\eta=\varepsilon/\varepsilon_{0}$. $\mathbf{M}$ is governed by the LLG
equation,
\begin{equation}
\partial_{t}\mathbf{M}=-\gamma \mu_0 \mathbf{M}\times\mathbf{H}_{\mathrm{eff}}%
+\frac{\alpha}{M_{\mathrm{s}}}\mathbf{M}\times\partial_{t}\mathbf{M},
\label{Eq-LLG}%
\end{equation}
where $\gamma,\alpha$ are the gyromagnetic ratio and Gilbert damping constant,
respectively. The effective magnetic field $\mathbf{H}_{\mathrm{eff}}%
=H\hat{z}+\mathbf{H}_{\mathrm{ex}}+\mathbf{h}$, consists of external,
exchange, and rf magnetic fields, where, the exchange field $\mathbf{H}%
_{\mathrm{ex}}=J\nabla^{2}\mathbf{m}$ with exchange constant $J$. For wave
vector $\mathbf{k}=k\hat{x}$, the coupled Eqs. (\ref{Eq-Maxs}) and
(\ref{Eq-LLG}) become,
\begin{equation}
\left(
\begin{array}
[c]{cc}%
(1+u_{k})k_{\varepsilon}^{2} & -iv_{k}k_{\varepsilon}^{2}\\
iv_{k}k_{\varepsilon}^{2} & (1+u_{k})k_{\varepsilon}^{2}-k^{2}%
\end{array}
\right)  \left(
\begin{array}
[c]{c}%
h_{x}\\
h_{y}%
\end{array}
\right)  =0. \label{Eq-H}%
\end{equation}

with $\omega_{\mathrm{M}}=\gamma \mu_0 M_{\mathrm{s}},$ $\omega_{\mathrm{H}}=\gamma \mu_0
H,$ $\omega_{k}=\omega_{\mathrm{H}}+J\omega_{\mathrm{M}}k^{2}-i\alpha\omega$
and
\begin{equation}
u_{k}=\frac{\omega_{k}\omega_{\mathrm{M}}}{\omega_{k}^{2}-\omega^{2}}%
,\;v_{k}=\frac{\omega\omega_{\mathrm{M}}}{\omega_{k}^{2}-\omega^{2}}.
\end{equation}

The secular equation of Eq. (\ref{Eq-H}) gives the dispersion relation for the
coupled microwave and spin wave modes or magnon-polaritons
\cite{Weiner1972, Gerson1974, Ruppin1987}
\begin{equation}
(1+u_{k})k^{2}=\left[  (1+u_{k})^{2}-v_{k}^{2}\right]  k_{\varepsilon}^{2}.
\label{Eq-Dis}%
\end{equation}

\begin{center}
\textbf{III. RESULTS}
\end{center}

\begin{center}
\textbf{A. Paramagnet ($J=0$) }
\end{center}

We first consider the simplest case of a paramagnet with uncoupled spins
$(J=0)$, which is equivalent with the macrospin model for unpinned
ferromagnetic order. $u_{k}=u,v_{k}=v$ are $k$ independent and
$k=k_{\varepsilon}\sqrt{1+u-v^{2}/(1+u)}$ for a given frequency $\omega$.
$h_{x}=-m_{x}$ is the dipolar field. The susceptibility $\chi=\partial m_{y}/\partial h_{y}$
resonates at $\omega_{\mathrm{FMR}}=\sqrt{\omega_{\mathrm{H}}(\omega
_{\mathrm{H}}+\omega_{\mathrm{M}})}$ with linewidth $\Delta\omega
_{\mathrm{FMR}}\simeq\alpha(2\omega_{\mathrm{H}}+\omega_{\mathrm{M}})$.
Rewriting the $h_{y}(x,t)=\psi(x)e^{-i\omega t}$, the potentials $\psi(x)$ in
the five separated regimes marked in Fig.\ref{Fig1}(b) read
\begin{subequations}
\begin{equation}
\psi_{1}(x)=e^{iqx}+Fe^{-iqx},\;\psi_{2}(x)=a_{1}e^{iqx}+a_{2}e^{-iqx},
\end{equation}%
\begin{equation}
\psi_{3}(x)=b_{1}e^{ikx}+b_{2}e^{-ikx},\;\psi_{4}(x)=c_{1}e^{iqx}%
+c_{2}e^{-iqx},
\end{equation}%
\begin{equation}
\psi_{5}(x)=Se^{iqx}.
\end{equation}
The coefficients $\{S,F,a_{1},a_{2},b_{1},b_{2},c_{1},c_{2}\}$ are determined
by the electromagnetic boundary conditions of continuity and flux conservation
at each interface. The transmission coefficient is
\end{subequations}
\begin{equation}
S=\frac{\left(  1-\beta^{2}\right)  t_{c}^{2}e^{i(k-q)d}}{\left(  1-\beta
r_{c}e^{i\phi}\right)  ^{2}-e^{2ikd}\left(  \beta-r_{c}e^{i\phi}\right)  ^{2}%
},\label{Eq-S}%
\end{equation}
where $\phi=q(L-d)$, $\beta=(\eta q-k)/(\eta q+k)$, introducing the scattering
coefficients of a single cavity wall $t_{c}=i/(i+q\ell)\ $and $r_{c}%
=-q\ell/(i+q\ell)$. We first inspect the resonant cavity modes identified by
the maxima of the transmission probability $|S|^{2}$ for non-magnetic loads
at
\begin{equation}
\left(  1+|r_{c}|^{2}\right)  \beta\sin(kd)=|r_{c}|\left[  \beta^{2}%
\sin(kd-\phi^{\ast})+\sin(kd+\phi^{\ast})\right]  ,\label{Eq-Res}%
\end{equation}
where $\phi^{\ast}=\phi+\mathrm{Arg}(r_{c})$. For $d=0$, we recover the
resonance condition of an empty cavity: $\phi_{n}^{\ast}=(n+1)\pi$, with mode
index $n=1,2,...$. It follows from Eqs. (\ref{Eq-Dis}) and (\ref{Eq-Res}) that
the resonance frequencies $\omega_{c,n}$ depend on both loading fraction $d/L$
and dielectric constant $\eta$. The cavity mode frequencies for a nonmagnetic
load are shown in Fig. \ref{Fig2}(a). Odd modes $\omega_{c,2j-1}$ have nodes
of the electric field at the sample position and depend only weakly on the
film thickness, in contrast to the even modes $\omega_{c,2j}$ with antinodes
that lead to redshifts. The anti-crossings of the cavity modes indicate
hybridization induced by the dielectric load that modulates its intrinsic
properties. The mode shifting due to the dielectric loading predicted here
is absent in the TC model. To avoid this complication, we focus our discussions on the nearly
empty cavity regime with loading rates $d/L<5\%$ and on odd cavity modes.

In the limit of long wavelength, i.e., $k\ll1/d$ only the leading term up to
order $k^{2}$ contributes. The transmission coefficient then reduces to
\begin{equation}
S_{n}=\frac{\kappa_{c,n}}{i(\omega-\omega_{c,n})-\kappa_{c,n}-ig_{n}%
^{2}\left(  \omega-\omega_{\mathrm{FMR}}+i\kappa_{s,n}\right)  ^{-1}},
\end{equation}
where $\kappa_{c,n}\simeq c^{3}/[{\small 2(L-d)\omega_{c,n}^{2}\ell^{2}]}$ is
the loss rate of the loaded cavity, and $\kappa_{s,n}\simeq\alpha
{\small \sqrt{\omega_{\mathrm{M}}^{2}+4\omega_{c,n}^{2}}/2}$ is that of the
magnetic film to the leading order in the Gilbert damping $\alpha$.

The effective coupling strengths $g_{n}$ depend on the parity of the cavity
modes, i.e., the odd-mode coupling scales as $\sqrt{d}$
\begin{subequations}
\begin{equation}\label{Eq-geff1}%
g_{2j-1}^{2}=\frac{d\omega_{\mathrm{M}}\left(  \omega_{\mathrm{M}}%
+\omega_{\mathrm{H}}\right)  }{2(L-d)}\cos^{2}\frac{\phi_{2j-1}^{\ast}}{2},
\end{equation}
while for even modes higher order corrections have to be included
\begin{align}
\label{Eq-geff2}g_{2j}^{2}  &  =\frac{d\omega_{\mathrm{M}}\left(
\omega_{\mathrm{M}}+\omega_{\mathrm{H}}\right)  }{2(L-d)}\cos^{2}\frac
{\phi_{2j}^{\ast}}{2}\\
&  \times\left\vert 1-d\eta q_{2j}\tan\frac{\phi_{2j}^{\ast}}{2}+\frac{\left(
d\eta q_{2j}\tan\left(  \phi_{2j}^{\ast}/2\right)  \right)  ^{2}}%
{6}\right\vert ,\nonumber
\end{align}
where $\phi_{n}^{\ast}$ is the phase at resonance frequency $\omega_{c,n}$.
Both odd and even modes can be tuned by the total number of spins $\propto d$ and
by the dielectric constant $\eta$. Anti-crossings between magnetic and cavity
modes occur at $\omega_{\mathrm{FMR}}=\omega_{c,n}$ or $\mu_0H_{\mathrm{res,n}
}=(-\omega_{\mathrm{M}}+\sqrt{\omega_{\mathrm{M}}^{2}+4\omega_{c,n}^{2}%
})/\left(  2\gamma\right) $. When not stated otherwise, we use the parameters
for YIG, with $\eta=15$ \cite{Sadhana2009}, $\gamma/(2\pi)=28\,\mathrm{GHz/T}$ and
$\mu_{0}M_{\mathrm{s}}=175\,\mathrm{mT}$ \cite{Manuilov2009}, while reported
$\alpha$'s range from $10^{-5}\sim10^{-3}$ \cite{Kajiwara2010, Heinrich2011,
Kurebayashi2011}. The resonance frequency $\omega_{c}$ and loss rate
$\kappa_{c}$ of the cavity is governed by its width $L$ and opacity $\ell$. We
choose $L=46\,\mathrm{mm}$ to be much larger than the film thickness $d$ and
the $n=3$ cavity mode (around $10\,\mathrm{GHz}$) as well as a $\kappa_{c,3}$
of the order of MHz, both of which can be tuned by $\ell$.
\begin{figure}[tbh]
\centering
\includegraphics[width=8.5cm]{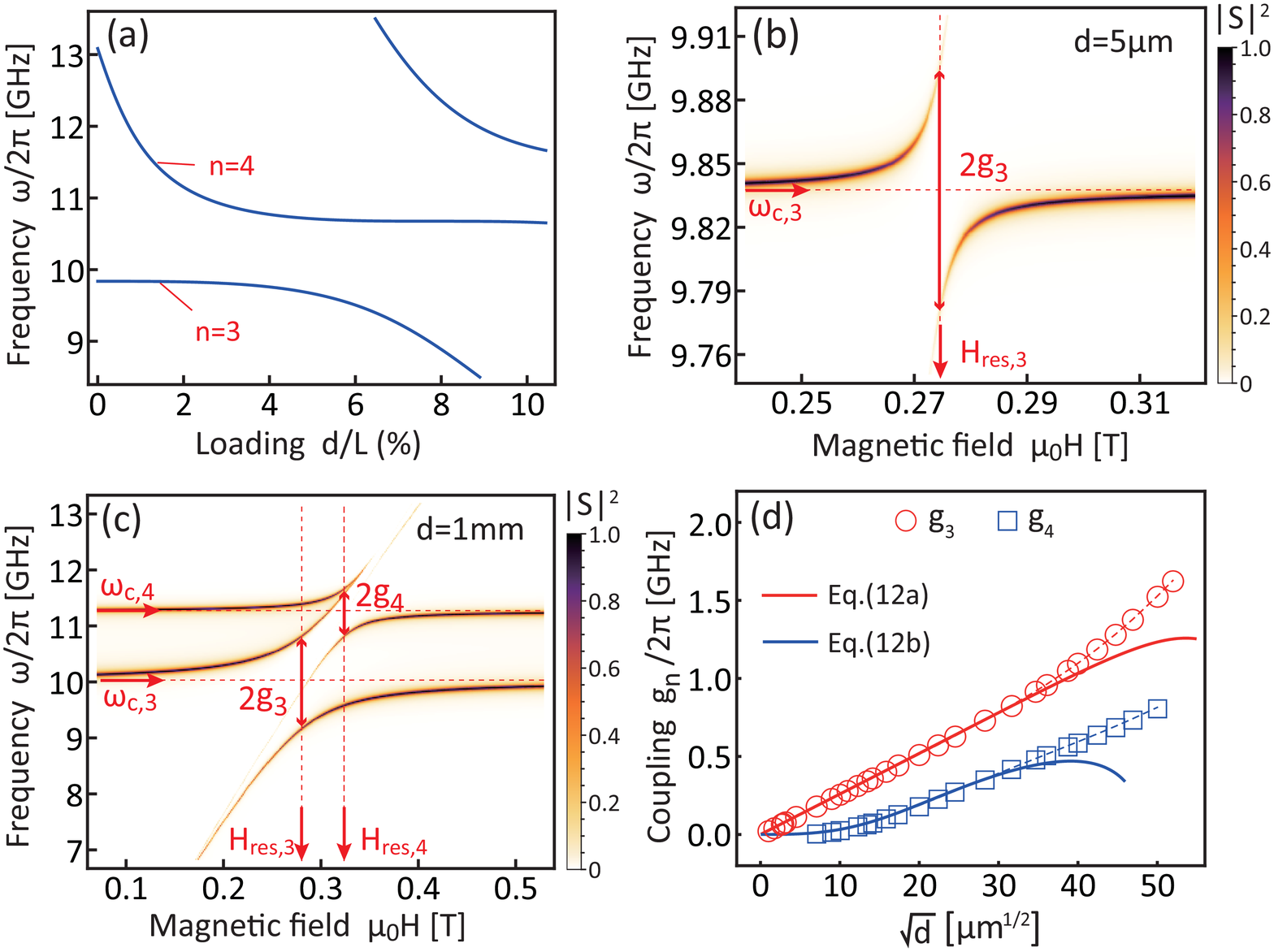}\caption{(a) Hybridized cavity eigen-modes [solutions of Eq. (\ref{Eq-Res})] in
the presence of a non-magnetic load as a function of loading rate with
dielectric constant $\eta=15$. Transmission spectra as a
function of magnetic field and frequency for two different magnetic films with
parameters (b) $d=5\;\mathrm{\muup m}$, and (c) $d=1$ mm.
(d) Thickness dependence of coupling strength for the third and
fourth modes. In the calculations, the length of the cavity $L=46$ mm,
cavity opacity $\ell/L=2$ except for 0.4 used in (c) to demonstrate the USC
with enough resolution, Gilbert damping $\alpha=3\times10^{-4}$, and exchange constant $J=0$
(paramagnetic limit).}%
\label{Fig2}%
\end{figure}

The transmission spectrum in the paramagnetic limit $J=0$ is shown for a thin
film with $d=5\,\mathrm{\mu m}\,(d/L=0.01\%)$ in Fig. \ref{Fig2}(b). At the resonant photon
frequency $\omega_{c,3}=9.84\,\mathrm{GHz}$, a coupling strength of
$g_{3}=57.77\,\mathrm{MHz}$ is extracted from the anti-crossing, where $g_{3}$
is much larger than both $\kappa_{c,3}=1.44\,\mathrm{MHz}$ and $\kappa
_{s,3}=3.04\,\mathrm{MHz}$, which implies strong coupling for a quasi 1D model
assuming homogeneous crossing section. However, when
$d=1\,\mathrm{mm}\; (d/L=2.17\%)$ in Fig. \ref{Fig2}(c), an additional anti-crossing
resonance at $\omega_{c,4}=11.27\,\mathrm{GHz}$ is observed with coupling
strength $g_{4}=0.43\,\mathrm{GHz}$. The main resonance for $\omega
_{c,3}=10.03\,\mathrm{GHz}$ has a coupling strength $g_{3}=0.83\,\mathrm{GHz}%
$, corresponding to a cooperativity $C=g_{3}^{2}/(\kappa_{c}\kappa_{s})=15072$
at loss rates $\kappa_{c,3}=34.71\,\mathrm{MHz}$ and $\kappa_{s,3}%
=3.10\,\mathrm{MHz}$, thereby approaching the USC regime of $g_{n}%
\gtrsim0.1\omega_{c,n}$. The coupling can also go into the
magnetically-induced transparency and Purcell effect regimes \cite{Zhang2014}
by tuning the parameters (not shown here).

The coupling strengths increase with $\sqrt{d}$ as shown in Fig.
\ref{Fig2}(d), where the red circles and blue squares are extracted from
numerical results for the full model calculations Eq. (\ref{Eq-S}), and the solid lines are the
analytical Eqs. (\ref{Eq-geff1}) and (\ref{Eq-geff2}) without any fitting
parameter. In the paramagnetic limit, the full model converges to Eq.
(\ref{Eq-S21}) when $kd\ll1$. The formula for $g_{n}$ begins to deviate when
$kd\simeq1$, where film thickness $d\simeq c/(\sqrt{\eta}\omega)=1.3\,$mm for
$\omega/2\pi=10$ GHz as shown in Fig. \ref{Fig2}(d).
Finite temperature can significantly reduce the spin polarization of paramagnets,
while ferromagnets are much more robust.

\begin{center}
\textbf{B. Ferromagnet ($J>0$) }
\end{center}

Now we consider finite exchange coupling, i.e., $J>0$. Equation (\ref{Eq-Dis})
has then 3 solutions for a given frequency and $\psi_{3}(x)$ is modified as
\end{subequations}
\begin{equation}
\psi_{3}(x)=\sum_{j=1}^{3}\left(  b_{1,j}e^{ik_{j}x}+b_{2,j}e^{-ik_{j}%
x}\right)  .
\end{equation}

\begin{figure}[tbh]
\centering
\includegraphics[width=8.5cm]{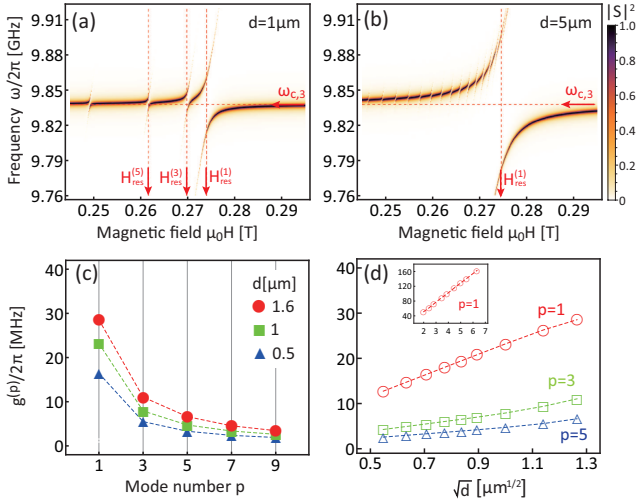}\newline\caption{ (a,b) : Transmission
for $d=1\;\mathrm{\muup m}$ and $d=5\;\mathrm{\muup m}$; (c,d): Mode-dependent
coupling strengths. In the calculations we used cavity opacity $\ell/L=2$,
Gilbert damping $\alpha=10^{-5}$, and ferromagnetic exchange constant
$J=3\times10^{-16}\;\mathrm{m^{2}}$ \cite{Serga2010}.}%
\label{Fig3}%
\end{figure}The magnetization dynamics now becomes sensitive to the surface
boundary conditions. Kittel \cite{Kittel1958} has shown that pinning of the
magnetization at the surface is required for SWR
(the absorption of spatially homogeneous microwaves by higher order spin waves),
and the symmetrically pinned boundaries merely render odd modes observable.
Here we adopt boundary conditions $\mathbf{m}(\left( L\pm
d\right)  /2)=0$, which can be justified by sufficiently strong surface
anisotropies \cite{Henryk1979,Liu2007}. The standing spin wave frequencies are
$\omega_{\mathrm{SWR}}^{\left(  p\right)  }=\sqrt{(\omega_{\mathrm{H}%
}+2J\omega_{\mathrm{M}}(p\pi/d)^{2})(\omega_{\mathrm{M}}+\omega_{\mathrm{H}%
}+2J\omega_{\mathrm{M}}(p\pi/d)^{2})}$ where $p\in%
\mathbb{N}
_{0}$. We consider in the following magnetic film thicknesses in the range
$0.1\sim5\,\mathrm{\mu m}$. Naively, exchange effects are appreciable when the
magnetic film thickness is comparable with the exchange length, $\lambda
_{\mathrm{ex}}\simeq17\,\mathrm{nm}$ for YIG, but they play a significant role
in the spectra of much thicker samples. For high quality magnetization
dynamics corresponding to a Gilbert damping $\alpha=10^{-5}$, the strong
coupling of the odd spin wave modes becomes evident from the transmission
spectrum for $d=1\,\mathrm{\mu m}\gg\lambda_{\mathrm{ex}}$. In Fig.
\ref{Fig3}(a), anti-crossings occurs at $\omega_{\mathrm{SWR}}^{\left(
p\right)  }$ with odd $p$ that are marked by red dashed lines at the SWR
magnetic fields $\mu_0 H_{\mathrm{res}}^{(p)}\simeq\left(  -\omega_{\mathrm{M}%
}-2J\omega_{\mathrm{M}}(p\pi/d)^{2}+{\small \sqrt{\omega_{\mathrm{M}}%
^{2}+4\omega_{\mathrm{c,3}}^{2}}}\right)  /(2\gamma)$.
The satellite anti-crossings are absent in the TC model.

In Fig. \ref{Fig3}(b), for $d=5\,\mathrm{\mu m}$, the anti-crossing resonances
of the lower spin wave modes condensate to the FMR splitting area. The
coupling strengths decrease with increasing mode number as shown in Fig.
\ref{Fig3}(c). The magnon-photon coupling for the main $p=1$ mode is
proportional to the total magnetization,
the coupling strength for spin waves $g^{(p)}$ $\propto \sqrt{d}/p$ for pinned surface magnetizations,
as shown in Fig. \ref{Fig3}(d).
For very thick films, i.e., $d>2\,\mathrm{\mu m}$, the spin wave modes start to overlap and
are difficult to distinguish. This collapse heralds the transition to the
paramagnetic macrospin model in spite of the surface pinning. The lowest
spin-wave mode is always dominant with $\sqrt{d}$-scaling that is not affected
by the transition, as shown in the inset of Fig. \ref{Fig3}(d).

\begin{center}
\textbf{C. Spin pumping }
\end{center}

Spin pumping detected by the ISHE is a useful
electrical technique to study magnetization dynamics \cite{Sandweg2010}. Let
us consider an ultrathin Pt film attached to the edge of the YIG slab as in
Fig. \ref{Fig1}(a). We assume free boundary conditions at the edges $y=0.$ The
magnetization dynamics at the interface then injects a spin current into the
Pt film that generates a Hall voltage $V_{\mathrm{ISHE}}%
=D_{\text{\textrm{ISHE}}}j_{\mathrm{s}}^{\mathrm{sp}}$ over the Pt wire, with
$D_{\mathrm{ISHE}}\equiv(2e/\hbar)\theta\xi(d/\sigma d_{y})\tanh(d_{y}/2\xi)$.
We illustrate strong coupling in the $V_{\text{\textrm{ISHE}}}$ spectrum here
for the paramagnetic (unpinned macrospin) limit $J=0$. The pumped spin current
can be written
\begin{equation}
j_{\mathrm{s}}^{\mathrm{sp}}=\frac{\hbar g_{r}^{\uparrow\downarrow}\omega
}{4\pi dM_{\mathrm{s}}^{2}}\mathrm{Im}\left[  \left(  u-\frac{v^{2}}%
{1+u}\right)  \frac{iv^{\ast}}{1+u^{\ast}}\right]  \int_{\frac{L-d}{2}}%
^{\frac{L+d}{2}}\mathrm{d}x|\psi_{3}(x)|^{2}.
\end{equation}
We assume that the Pt wire has the width $d_{y}=10\,\mathrm{nm}$ with
conductivity $\sigma=10^{7}\,\mathrm{(m\cdot\Omega)}^{-1}$, spin mixing
conductance $g_{r}^{\uparrow\downarrow}=10^{19}\,\mathrm{m}^{-2}$, spin Hall
angle $\theta=0.11$ and spin diffusion length $\xi=1.5\,\mathrm{nm}$
\cite{Weiler2013}. The spin back-flow contributes a minor correction that we
disregard since $\xi\ll d_{y}$. The rf magnetic amplitude is chosen as
$\mu_{0}h_{0}=10\,\mathrm{\mu T}$. The microwave power absorption is defined
as the integration of the Poynting vector enclosed by a section of the volume
of the sample reads,
\begin{equation}
P_{\mathrm{abs}}=\frac{\mu_{0}d_{y}d_{z}\omega}{2}\mathrm{Im}\left(
u-\frac{v^{2}}{1+u}\right)  \int_{\frac{L-d}{2}}^{\frac{L+d}{2}}%
\mathrm{d}x|\psi_{3}(x)|^{2}.
\end{equation}
\begin{figure}[tbh]
\centering
\includegraphics[width=7cm]{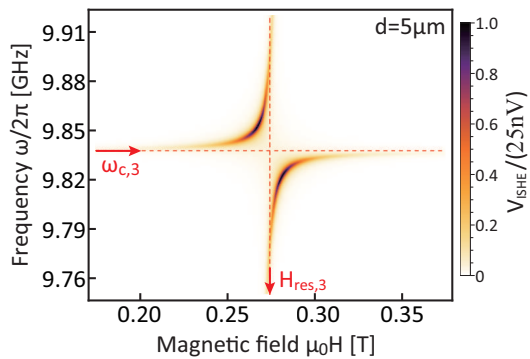}\newline\caption{Inverse spin Hall
voltage spectrum. For a cavity $\ell/L=2$, Gilbert damping $\alpha
=2\times10^{-3}$, and $J=0$ (paramagnetic limit).}%
\label{Fig4}%
\end{figure}

By substituting $u$ and $v$, we find that $j_{\mathrm{s}%
}^{\mathrm{sp}}/P_{\mathrm{abs}}\propto\omega_{\mathrm{M}}(\omega_{\mathrm{M}%
}+\omega_{\mathrm{H}})/\alpha\lbrack\omega^{2}+(\omega_{\mathrm{M}}%
+\omega_{\mathrm{H}})^{2}]$ is almost a constant near the resonance, which
proves that the spin pumping is a reliable measure of the microwave
absorption. $V_{\text{\textrm{ISHE}}}$ as a function of rf frequency and
magnetic field is shown in Fig. \ref{Fig4} for film thickness
$d=5\,\mathrm{\mu m}$. In the present symmetric configuration there are no
surface states that might interact strongly with the Pt contact
\cite{Sandweg2010, Kapelrud2013}. The calculations in the presence of exchange
(not shown) support our conclusions.

\begin{center}
\textbf{IV. CONCLUSIONS }
\end{center}

To summarize, we develop a scattering theory to study exchange magnon-polaritons, i.e.,
the hybridized magnetization and microwave dynamics, beyond the
paramagnetic/macrospin and RWA that are implicit in
the TC model. Our method and scattering coefficient Eq.
(\ref{Eq-S}) are valid for the full parameter range spanning the weak to
strong coupling limits. The conventional input-output formula Eq.
(\ref{Eq-S21}) is valid for odd cavity modes and only to leading order in the
film thickness $d$, otherwise the cavity properties are strongly modified by
the load. The exchange interaction between spins leads to strong coupling not
only for the FMR mode but also for standing spin waves. The magnon-photon
coupling depends on both the materials parameters and the spin wave mode
index, e.g., decrease with increasing mode number. We confirm the transition
from weak coupling, to strong coupling, to magnetically induced transparency
and to ultra-strong coupling regimes. Spin pumping from magnon-polaritons into
metallic thin film contacts shows pronounced anti-crossing spectra, which
allows electric readout of magnon-photon states. We believe that our results
will help to understand and engineer the coherent hybridization of
ferromagnetic and superconducting order parameters in microwave cavities
\cite{Tabuchi2014}.

\begin{center}
\textbf{V. ACKNOWLEDGMENTS }
\end{center}

We acknowledge helpful discussions with Yaroslav Blanter,
Johannes Lotze, Hannes Maier-Flaig, Babak Zare Rameshti and Ka Shen.
The research leading to these results has
received funding from the European Union Seventh Framework Programme
[FP7-People-2012-ITN] under grant agreement 316657 (SpinIcur). It was
supported by JSPS Grants-in-Aid for Scientific Research (Grant Nos. 25247056,
25220910, 26103006), FOM (Stichting voor Fundamenteel Onderzoek der
Materie),\ the ICC-IMR, EU-FET InSpin 612759, and DFG Priority Programme 1538
\textquotedblleft{Spin-Caloric Transport}\textquotedblright\ (BA 2954/1, GO 944/4)
and the collaborative research center SFB631 (C3).

\end{document}